\documentclass[12pt]{article}
\usepackage{authblk}
\usepackage{amsmath}
\usepackage{graphicx}
\usepackage{float}
\usepackage{xcolor}
\usepackage[margin=1.2in]{geometry}

\usepackage[light,math]{kurier}
\usepackage[T1]{fontenc}

\title{Systematic errors related to quadrupole misplacement in an all-electric storage ring for proton EDM experiment}
\author[1]{Sel{\c c}uk Hac{\i}{\"o}mero{\u g}lu}
\author[1,2]{Yannis K. Semertzidis}
\affil[1]{\small IBS, Center for Axion and Precision Physics, \textit{Daejeon, 34051, South Korea}}
\affil[2]{\small KAIST, Physics Department, \textit{Daejeon, 34141, South Korea}}

\begin{document}

\maketitle

\begin{abstract}
Misplacement of electrostatic elements can pose false EDM signal in a storage ring EDM experiment because of coupling between vertical electric field and magnetic dipole moment. A vertically misplaced quadrupole introduces electric field proportional to its misplacement, changing periodically in the particle's rest frame during the storage. This leads to accumulation of vertical spin component at every revolution. We investigated this effect by simulating a proton in an all-electric ring with several quadrupole scenarios. It turns out that the misplacement of quadrupoles is a critical item to keep under control, for which we propose several methods. These include tuning the frequency of the RF cavity, making use of additional correction quadrupoles and using quadrupoles with weaker focusing strength.
\end{abstract}

\section{Introduction}
A storage ring with all-electric elements can be used to probe the electric dipole moment (EDM) of protons \cite{ref:edm_in_sring, ref:edm_paper}. The electrostatic deflectors in such a ring both store the beams and couple with the EDM to induce a spin precession. The spin precession under electric and magnetic fields is governed by the T-BMT equation  \cite{ref:bmt}. With the additional EDM terms, it is given as.
\begin{equation}
\begin{split}
\frac{d \vec s}{dt} = \frac{e}{m} &\vec s \times \Bigg[ \Bigg(\left( G + \frac{1}{\gamma}\right) \vec B - \frac{\gamma G}{\gamma+1}\vec \beta (\vec \beta \cdot \vec B)-\Big(G+\frac{1}{\gamma+1} \Big)\frac{\vec \beta \times \vec E}{c} \Bigg) \\
& + \frac{\eta}{2c}\Bigg(\vec E -\frac{\gamma}{\gamma+1} \vec \beta (\vec \beta \cdot \vec E) + c \vec \beta \times \vec B \Bigg) \Bigg]
\end{split}
\label{eq:bmt}
\end{equation}
where $c$, $e$ and $m$ are the speed of light, the electric charge and the mass of the particle, $G$ is the anomalous magnetic moment ($\approx 1.8$ for protons),  $\vec \beta$ and $\gamma$ are the relativistic velocity and the Lorentz factor, $\vec B$ and $\vec E$ are the magnetic and electric field vectors respectively. $\eta$ is the EDM coefficient, defined by $\vec d_p=\frac{\eta q \hbar}{2mc} \vec s$.

Neglecting the magnetic field it becomes
\begin{equation}
\frac{d \vec s}{dt} = \frac{e}{m} \vec s \times \Bigg[ -\Big(G+\frac{1}{\gamma+1} \Big)\frac{\vec \beta \times \vec E}{c} \\
 + \frac{\eta}{2c}\Bigg(\vec E -\frac{\gamma}{\gamma+1} \vec \beta (\vec \beta \cdot \vec E) \Bigg]
\label{eq:bmt_el}
\end{equation}
The storage ring EDM experiment \cite{ref:edm_paper} aims to measure the second term of the Equation, which is proportional to $\eta$ and the radial electric field $E_r$. That will induce a few nrad/s of  $s_y$ for $\eta \approx 2 \times 10^{-15}$, corresponding to EDM of $d_p \approx 10 ^{-29}$ e$\cdot$cm inside a ring with $<E_r> \approx 5$ MV/m. The other term in the equation is a potential systematic error source, being enhanced with strong vertical electric field $E_y$.

In addition, the spin grows a radial component under certain circumstances. Neglecting magnetic fields and transverse components of the velocity, the dominant term leading to horizontal spin precession comes from the third term of Equation \ref{eq:bmt}.
\begin{equation}
\frac{ds_r}{dt} = -\frac{e}{m} s_l \Bigg[  \Big(G+\frac{1}{\gamma+1} \Big)\frac{\beta_l E_r}{c} \Bigg]
\label{eq:bmt_wa_lab_frame}
\end{equation}
The angle between the spin and momentum can be calculated by subtracting from this term the angular momentum vector. In the particle's rest frame in horizontal plane, its rate of change approximates to
\begin{equation}
	\omega_a = -\frac{e}{m} \Big(G-\frac{1}{\gamma^2-1}\Big) \frac{\beta_l E_r}{c}
\label{eq:bmt_wa_rest_frame}
\end{equation}
Ideally it is possible to freeze the horizontal spin precession with respect to the velocity by injecting the beam with a specific energy, determined by $\gamma_0=\sqrt{1/G+1}$ from Equation \ref{eq:bmt_wa_rest_frame}. This is the basic idea of ``frozen spin method'' \cite{ref:edm_in_sring, ref:frozen_spin_2, ref:frozen_spin_3}. For protons, this condition can be satisfied with $p_0\approx 0.7$ GeV/c, coined as ``magic momentum''.

In practice there will be a momentum spread around the magic momentum. Besides, almost all particles make betatron oscillations due to transverse velocity and their transverse offset at the time of injection. Nevertheless, their momentum can be made to oscillate around the magic value by using an RF cavity, so that the average spin of the beam oscillates horizontally around the momentum vector \cite{ref:selcuk_rk4}. There is still a tiny drift coming from the uncorrected second order term. This tiny drift eventually makes the radial spin component to grow as much as a radian within the so-called ``spin coherence time''. As a consequence, spin decoherence comes with off-magic momentum.

The first term of Equation \ref{eq:bmt_el} shows that under a vertical electric field, the off-magic momentum particles will grow a vertical spin component as well. In the storage ring EDM experiment, this scenario can be encountered as a result of misalignment of electrostatic elements like deflectors, quadrupoles, etc \cite{ref:m_bai}. Note that, even though the average $<E_y>=0$, the effects are finite.

\section{Application to the pEDM experiment}

A recently published paper \cite{ref:edm_paper} describes a storage ring experiment for probing the EDM of proton at the $10^{-29} \text{ e}\cdot\text{cm}$ level. Longitudinally polarized  counter-rotating proton beams will be injected at approximately 0.7 GeV/c and stored for 1000 seconds inside a 500m long ring. The ring is composed of electrostatic elements with a simple FODO lattice. The momentum of the beams will be averaged to the magic value by an RF cavity as explained above. The radial electric field will couple with the EDM to grow a vertical spin component. The beams will be continuously extracted to the polarimeter \cite{ref:polarimeter} for spin measurement during storage. The total measurement time will be of order of $10^7$ seconds. The quadrupoles are designed to be 40cm long with roughly 35 MV/m$^2$ focusing strength.

\section{Misplacement of quadrupoles}

In case of misplacement of focusing elements, the closed orbit can be shifted both horizontally and vertically. The vertical offset causes a net vertical electric field locally in the ring, which causes off-plane spin precession similar to the EDM signal according to the first term of Equation \ref{eq:bmt_el}. It also causes an off-magic momentum because of the vertical motion it introduces. Similarly, the horizontal offset also causes a change in momentum even if it was magic at the time of injection. Combining those two effects enhances the false EDM signal. 

We made simulations to study how the spin precession is influenced by the misplacement of quadrupoles. The simulations were made with a tracking code based on fourth order Runge-Kutta integrator to simulate one proton  inside alternating gradient all-electric lattice. The details of the simulation tool are descibed in \cite{ref:selcuk_rk4} for a weak focusing all-electric and \cite{ref:benchmarking_paper} for a weak focusing magnetic ring. 

We studied the false EDM signal originating from basically two scenarios: random and symmetric misplacement of quadrupoles. While $\eta$ is kept zero in both cases, the vertical spin component grows much faster than nrad/s rate. On the other hand, this effect can be suppressed by using several methods, namely RF frequency tuning, using correction quadrupoles and beam-based quadrupole alignment.


\subsection{Random misplacement of quadrupoles}
\begin{figure}
	\centering
	\includegraphics[width=0.7\linewidth]{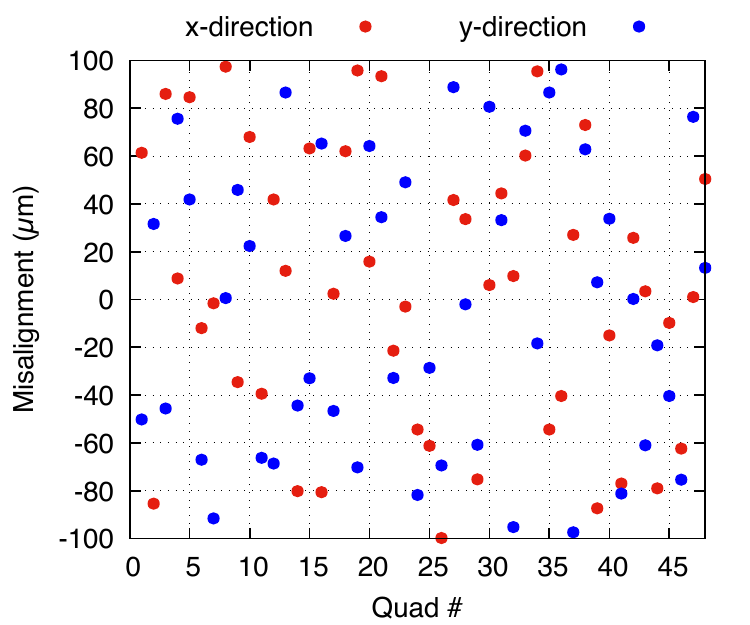}
	\caption{The quadrupoles are misaligned between $\pm 100 \mu$m both horizontally and vertically.}
	\label{fig:random_quad_misalignments}
\end{figure}

\begin{figure}
	\centering
	\includegraphics[width=0.7\linewidth]{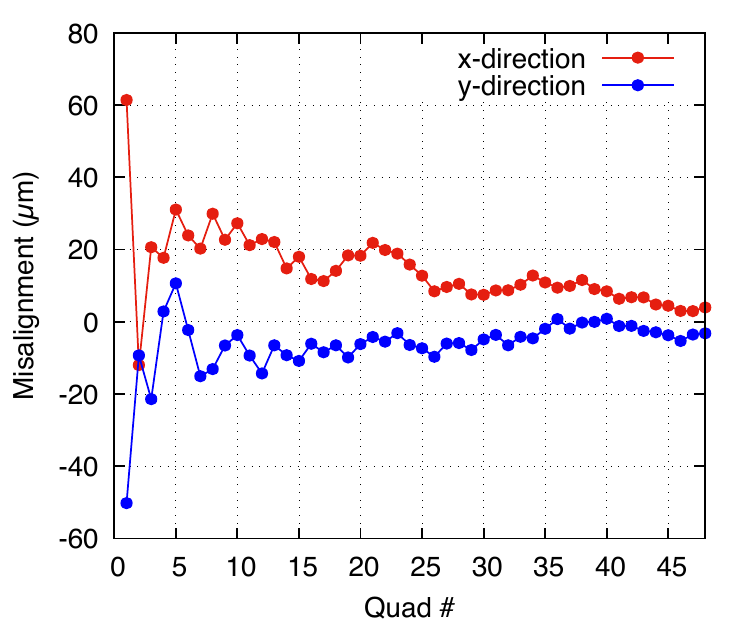}
	\caption{Passing through each quadrupole, the particle sees a net misalignment along the ring.}
	\label{fig:random_misalignments_avg}
\end{figure}

In these simulations we misplaced quadrupoles in horizontal and vertical directions separately in a random fashion. Keeping the pattern the same, the misplacements were scaled at various simulations. Figure \ref{fig:random_quad_misalignments} shows the offset of each quadrupole in the simulations for the case of $<100 \mu$m maximum. While traveling around the ring, the particle sees a net offset of a few $\mu$m on average (Figure \ref{fig:random_misalignments_avg}), as the misaligned quadrupoles do not necessarily average to zero. This causes a distortion of the closed orbit.

\begin{figure}
	\centering
	\includegraphics[width=0.7\linewidth]{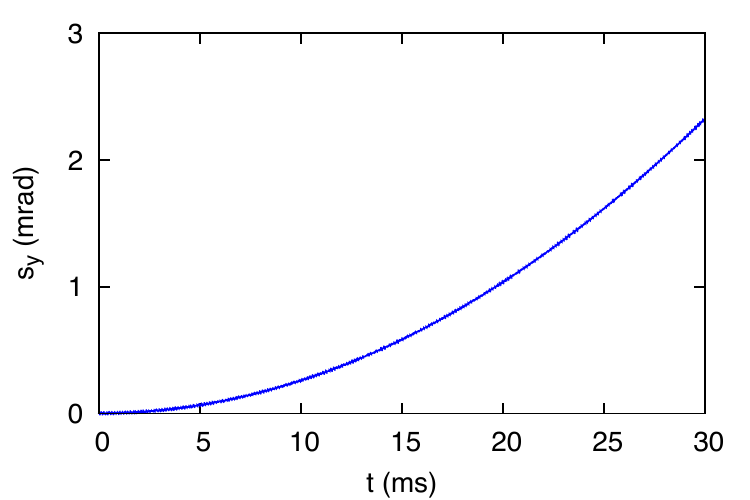}
	\caption{Random quadrupole misalignment makes $s_y$ grow quadratically overweighting the EDM signal.}
	\label{fig:sy_no_rf_correction}
\end{figure}

\begin{figure}
	\centering
	\includegraphics[width=0.7\linewidth]{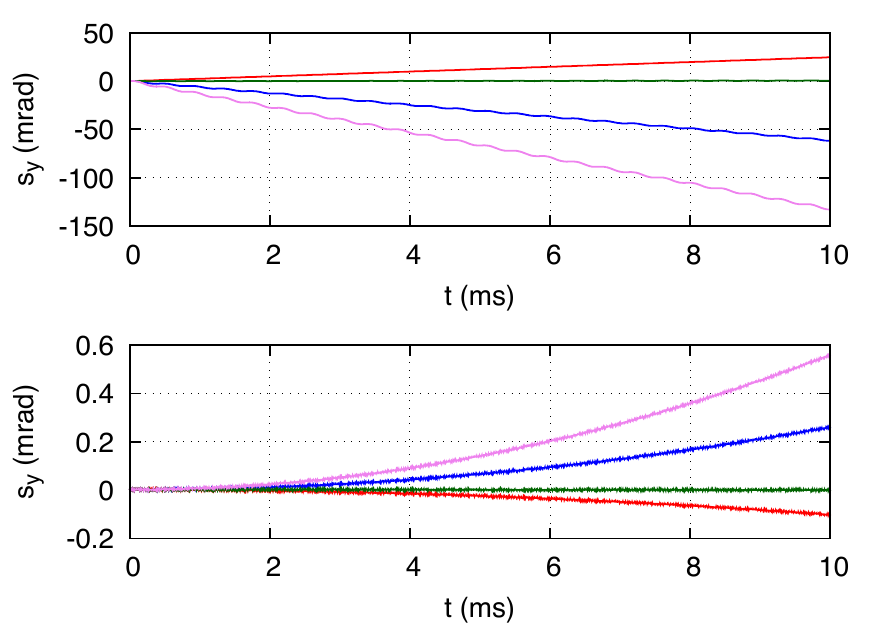}
	\caption{Precession of $s_y$ can be controlled by tuning the RF cavity. Each color corresponds to a specific RF frequency. There is a specific frequency which minimizes both $s_r$ and $s_y$ at the same time.}
	\label{fig:sy_with_rf}
\end{figure}

Figure \ref{fig:sy_no_rf_correction} shows the vertical spin component growing quadratically due to these misplacements. Note that this effect is enhanced by two factors: The presence of vertical electric fields and off-magic momentum. The latter also causes accumulation of radial spin component as shown in Equation \ref{eq:bmt_wa_rest_frame}. One can tune the frequency of the RF cavity to adjust the momentum of the particle and minimize both $s_r$ and $s_y$. Figure \ref{fig:sy_with_rf} shows the effect of RF tuning on $s_y$. There is a specific RF frequency which stops the drift of the horizontal spin precession and when the spin is aligned with the velocity direction, $s_y$ freezes too. This correction addresses the problem of distortion of the orbit in general, not limited to the misalignment of the quadrupoles. Therefore, while not studied in particular, we expect this method to fix the effect of misalignment of all electrostatic elements like deflectors, sextupoles, etc.

As an alternative solution, one can use additional weaker quadrupoles for correcting the false EDM signal. We simulated a particle with longitudinally aligned spin ($\phi=0$) and optimized the horizontal position of the correction quadrupole to 2.192 cm which gives the smallest radial spin precession rate. Then, we set the horizontal position of the quadrupole, simulated the particle with $\phi=45^0$ and minimized the vertical spin precession rate (false EDM signal) by changing the vertical position of the quadrupole. Note that the false EDM signal maximizes at $\phi=90^0$ and the real EDM signal minimizes with bigger $\phi$ according to Equation \ref{eq:bmt_el}. For these calculations $45^0$ and $90^0$ do not make a big difference because $\eta$ is zero.

\begin{figure}
	\centering
	\includegraphics[width=0.7\linewidth]{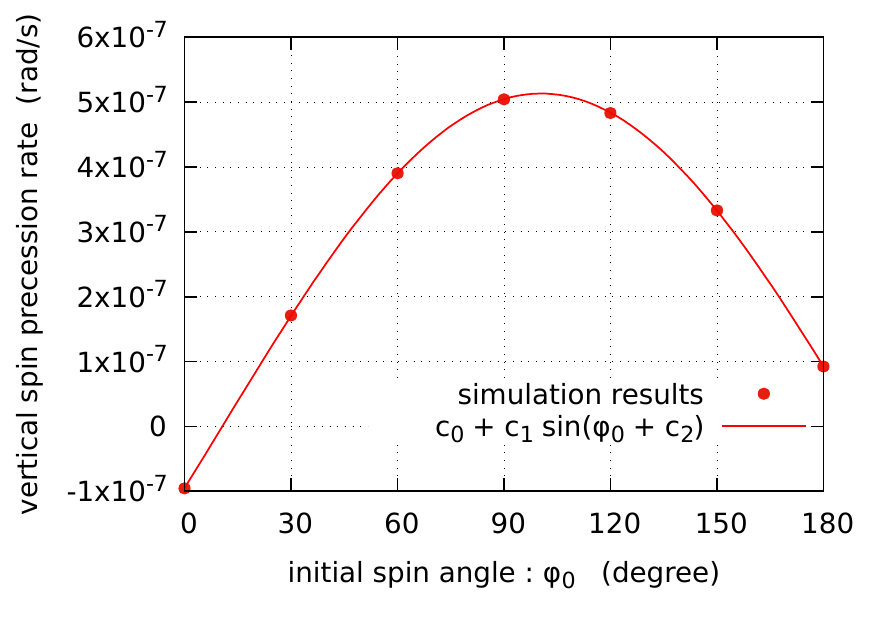}
	\caption{The dependence of vertical spin component on $\phi$. The coefficients fitting the data are: $c_0=8.6 \times 10^{-9}, c_1=5.1 \times 10^{-7}$ and $c_2=-11.5$ degrees.}
	\label{fig:sy_vs_angle_lebedev}
\end{figure}

Figure \ref{fig:sy_vs_angle_lebedev} shows the vertical spin value changing with $\phi$ after the correction of the quadrupoles. Note the large phase $c_2$ and the strong dependence on the spin angle ($c_1$). This means a small error in the spin polarization causes a large false EDM signal.

One solution to this issue is to use weaker quadrupoles. The lattice in the referenced paper \cite{ref:edm_paper} (Lattice-1) has four sets of quadrupoles with focusing strengths. We changed the strength of each quadrupole to achieve a stable storage with a much smaller value (Lattice-2). These values are chosen arbitrarily to give a stable storage. The horizontal and vertical emittance are 5 mm$\cdot$mrad and 0.4 mm$\cdot$mrad respectively. Table \ref{tbl:quad_strength} compares the strength of each quadrupole for the two lattices. 

\begin{table}
	\renewcommand*{\arraystretch}{1.3}
	\caption{Lattice-1 refers to the lattice described in \cite{ref:edm_paper}. Lattice-2 has everything the same except for the quadrupole strength.}
	\centering
	\begin{tabular}{c r r}
		\hline
		Quadrupole & Lattice-1 & Lattice-2 \\
		\hline
		$k_1$ (V/m$^2$) & $-3.4 \times 10^7$ & 0 \\
		$k_2$ (V/m$^2$) & $4.2 \times 10^7$ & $10^5$ \\
		$k_3$ (V/m$^2$) & $3.7 \times 10^7$ & $-2\times 10^5$ \\
		$k_4$ (V/m$^2$) & $-3.2 \times 10^7$ & 0 \\
		\hline
	\end{tabular}
	\label{tbl:quad_strength}
\end{table}

\begin{figure}
	\centering
	\includegraphics[width=0.7\linewidth]{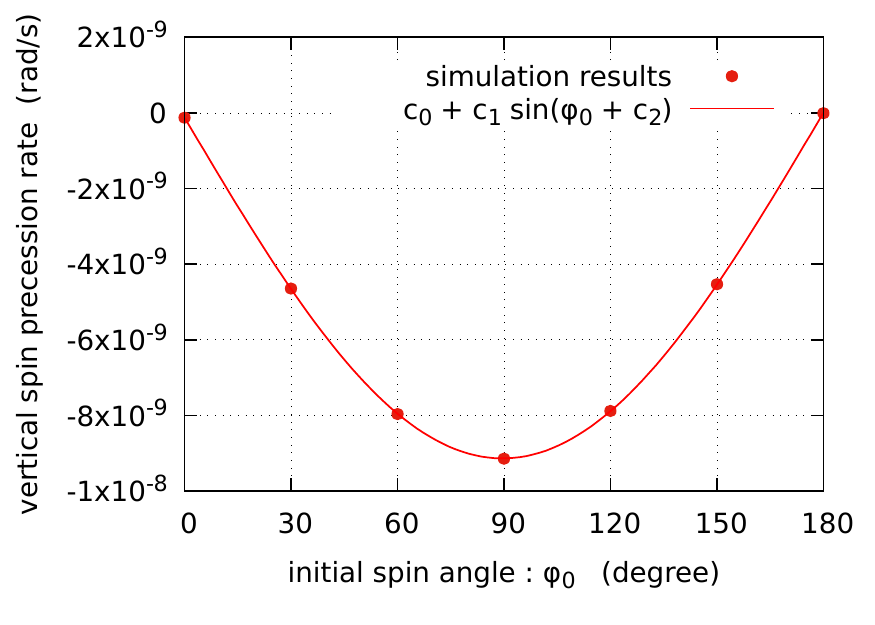}
	\caption{The false EDM signal becomes much smaller with weaker quadrupole strengths. The fit parameters turn out to be $c_0=-5.9 \times 10^{-11}, c_1=-9.1 \times 10^{-9}$ and $c_2=0.42$ degrees.}
	\label{fig:sy_vs_angle_talman}
\end{figure}

Figure \ref{fig:sy_vs_angle_talman} shows that all the fit parameters decrease considerably with weaker quadrupoles. For instance, $\phi_0=1\text{ mrad}\approx 0.06^0$ leads to about 0.1 nrad/s vertical spin precession, which is an order of magnitude less than the EDM signal.

\subsection{Geometric phase effect}
Geometric phase \cite{ref:berry_geom_phase}, \cite{ref:geom_phase_nEDM} appears in the presence of periodic distortions of field. The effect originates from the coupling of $s_y$ and $s_r$.  In such cases, $s_r$ is not symmetric while $s_y$ rises and falls. Therefore the amount of rise and fall differ. This can be visualized imagining Rubick's cube. One needs to follow the correct order when taking the move back, otherwise there is a residual effect. The repetition of this at each cycle causes an accumulation of the vertical spin component \cite{ref:y_orlov_geom_phase}. 

The spin accumulation in the case of random misplacement of quadrupoles originates from the geometric phase effect. Each misplaced quadrupole introduces an additional transverse electric field, changing periodically around the ring, potentially causing the geometrical phase effect.

\subsection{Symmetric distortions along the ring}
\begin{figure}
	\centering
	\includegraphics[width=0.55\linewidth]{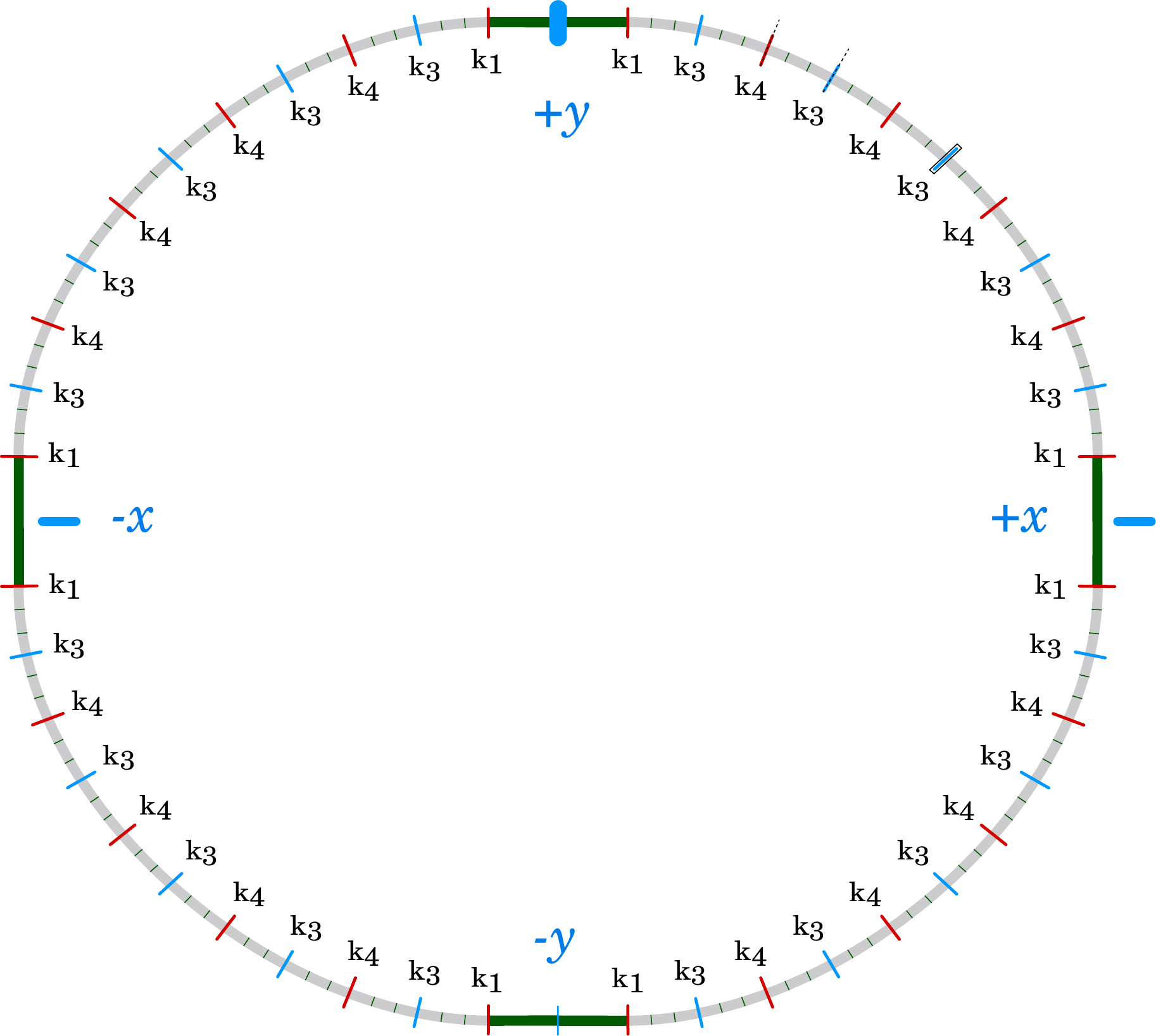}
	\caption{Geometric phase effect arises in some configurations with alternating field. $\pm x$ and $\pm y$ shows the direction of the misalignment of the quadrupoles. $x$ and $y$ stand for horizontal and vertical respectively.  The average field is zero along the ring, but the net effect on $s_y$ is nonzero due to the order of oscillations in perpendicular directions. Perpendicular directions do not have to be $90^0$ apart.}
	\label{fig:quad_misalignment}
\end{figure}

Figure \ref{fig:quad_misalignment} shows a marginal case of geometric phase effect in a storage ring with four misplaced quadrupoles at four ends of the ring. The misplacement was introduced in an alternating fashion to maximize the effect. This way, the particle experiences spin precession in perpendicular directions one after another in a consecutive fashion. Eventually, there is a residual amount of $s_y$ at each cycle because of the geometrical phase effect.

\begin{figure}
	\centering
	\includegraphics[width=0.7\linewidth]{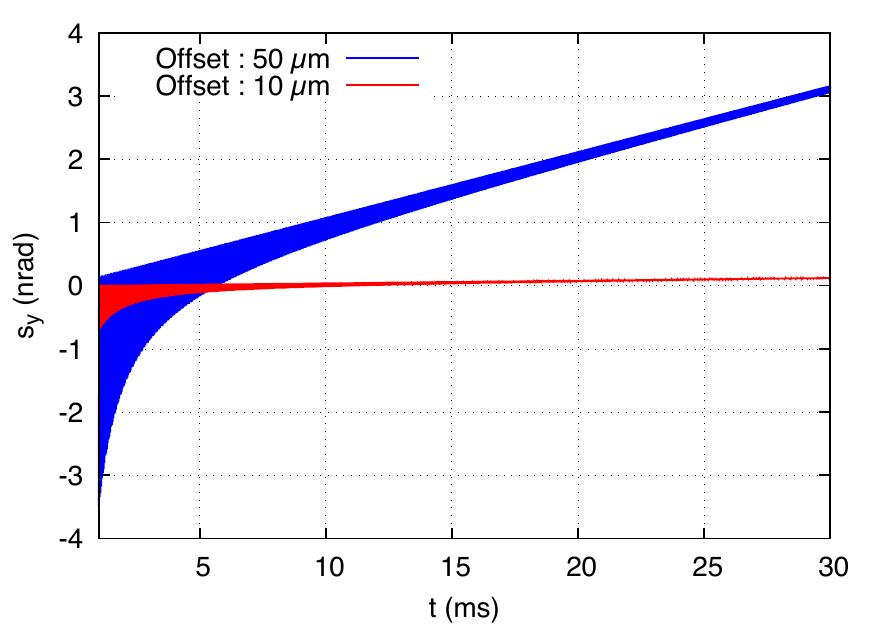}
	\caption{Even though the average quadrupole misalignment is zero along the ring, the geometric phase effect accumulates a relatively big spin component. In the case of $\pm50\mu$m offset, the spin precession rate is $\approx 110$ nrad/s and in the case of $\pm 10 \mu$m, it is $\approx 5$ nrad/s.}
	\label{fig:sy_geom_phase}
\end{figure}

Figure \ref{fig:sy_geom_phase} shows the running average of $s_y$ as obtained in simulations with the configuration shown in Figure \ref{fig:quad_misalignment}. The running average after $N$ time steps is defined as 
\begin{equation*}
	s_y^N=\frac{1}{N}\sum_{n=1}^{N-1}s_y^n
\end{equation*}
with $n$ representing each time step. In one of the cases the offset of each quadrupole is $\pm50 \mu$m, and in the other case it is $\pm10 \mu$m. Note the quadratic dependence of the accumulation of $s_y$ on quadrupole offset. This comes from the fact that both radial and vertical oscillations scale locally with the transverse fields, hence transverse misplacements.

As seen in the figure, the offset should be aligned to better than $10 \mu$m to suppress false EDM signal due to this effect.

\begin{figure}
	\centering
	\includegraphics[width=0.7\linewidth]{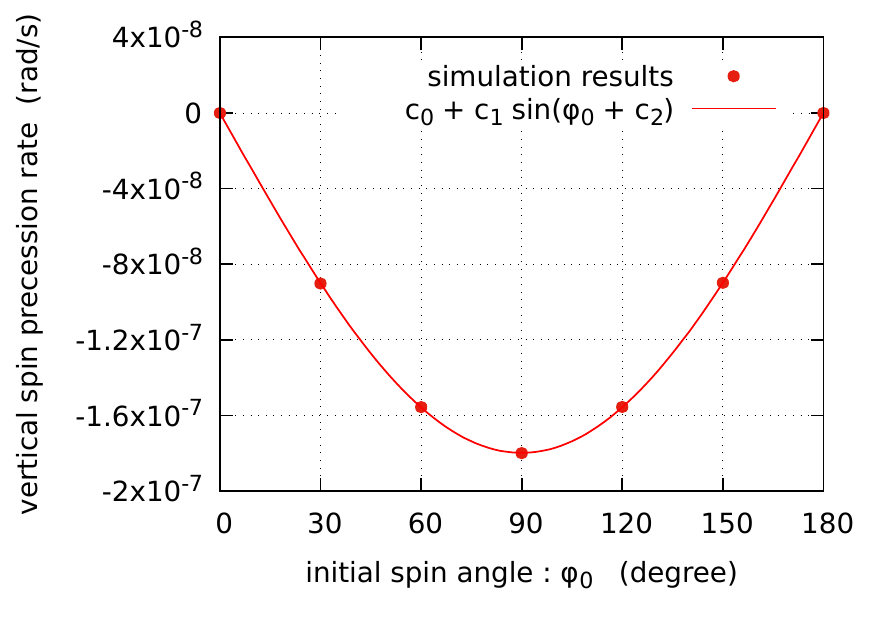}
	\caption{The simulation shows that the geometric phase effect becomes negligible with weaker quadrupoles. The fit parameters are: $c_0=-1.9 \times 10^{-11}$, $c_1=-1.8 \times 10^{-7}$, $c_2=0.03$ degree.}
	\label{fig:sy_vs_angle_talman_geom_phase}
\end{figure}

We repeated the same simulations in the lattice with weaker quadrupoles and an additional correction quadrupole as explained above. The regular quadrupoles are given misalignments of $10 \mu$m with the configuration shown in Figure \ref{fig:quad_misalignment}. Again, the false EDM signal became negligible compared to the real EDM signal after optimizing the position of the correction quadrupole (Figure \ref{fig:sy_vs_angle_talman_geom_phase}).

\subsection{Beam-based alignment}
\begin{figure}
	\centering
	\includegraphics[width=0.7\linewidth]{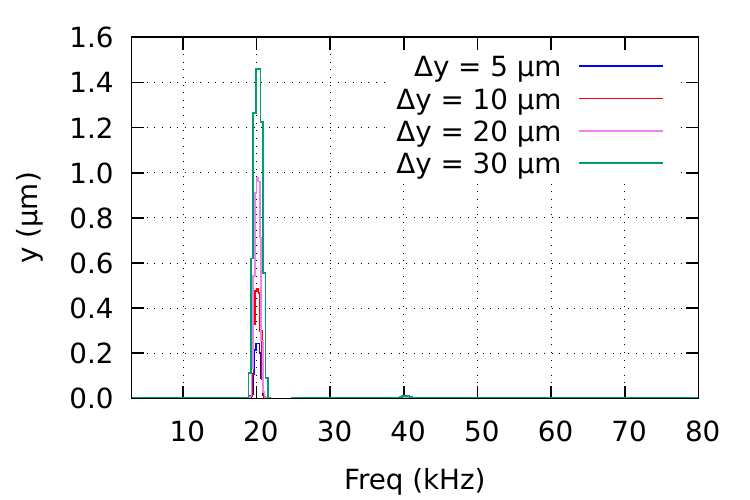}
	\caption{Vertical oscillation of the particle has a Fourier component at the modulation frequency of the quadrupole. The amplitude of the peak becomes smaller with smaller misalignment.}
	\label{fig:beam_based_alignment}
\end{figure}

It may be possible to align the quadrupoles by separately modulating them at specific frequencies. We made several simulations with a particle in a lattice with one vertically misplaced quadrupoles. The misplacement $\Delta y$ was set to various values between $5 \mu$m and $30 \mu$m. We modulated the strength of the misaligned quadrupole by about 3\% at 20 kHz. This modulation was seen in the vertical oscillation of the particle. FFT of the vertical position $y$ after 4 ms simulation shows peaks at the modulation frequency as seen in Figure \ref{fig:beam_based_alignment}. The amplitude of the peak is proportional to the misalignment. The method requires measuring the vertical position of the beam by a BPM with sub-$\mu$m resolution and then actuating the quadrupole in a way to minimize the peak. This way a few $\mu$m alignment can be achievable. 

\section{Conclusion}

This study investigates the effect of misplacement of quadrupoles on the spin of a proton inside an all-electric storage ring. Misalignment of quadrupoles is a source of a systematic error, mimicking the EDM signal. This mainly originates from the coupling between vertical electric field and the spin. 

We investigated several methods in simulations to solve this issue: namely tuning the frequency of the RF cavity, and positioning a correction quadrupole in the ring. While these methods help reducing the effect, lowering the quadrupole strength seems inevitable for a negligible false EDM signal.

The quadrupole misplacement should also be kept low for a good control of the effect. The simulations show that modulation of a misplaced quadrupole modulates the vertical oscillation of the particle as well. This feature can be exploited to minimize the misplacement of the quadrupoles. The vertical position should be measured by 0.2 $\mu$m to align the relative position of a quadrupole by 5 $\mu$m. 


\end{document}